\documentclass[12pt,preprint]{emulateapj}

\begin{document}
\title{Photometric Constraints on the Redshift of $z\sim10$ candidate
  UDFj-39546284 from deeper WFC3/IR+ACS+IRAC observations over the
  HUDF\altaffilmark{1}}

\author{R. J. Bouwens\altaffilmark{2,3}, 
P. A. Oesch\altaffilmark{3,\dag}, 
G. D. Illingworth\altaffilmark{3}, 
I. Labb{\'e}\altaffilmark{2}, 
P. G. van Dokkum\altaffilmark{4},
G. Brammer\altaffilmark{5},
D. Magee\altaffilmark{3}, 
L.R. Spitler\altaffilmark{7,8}, 
M. Franx\altaffilmark{2}, 
R. Smit\altaffilmark{2}, 
M. Trenti\altaffilmark{6}, 
V. Gonzalez\altaffilmark{3,9},
C. M. Carollo\altaffilmark{10}
}

\altaffiltext{2}{Leiden Observatory, Leiden University, NL-2300 RA Leiden, Netherlands}
\altaffiltext{3}{UCO/Lick Observatory, University of California, Santa Cruz, CA 95064}
\altaffiltext{4}{Department of Astronomy, Yale University, New Haven, CT 06520}
\altaffiltext{5}{European Southern Observatory, Alonso de C{\'o}́rdova 3107, Casilla 19001, Vitacura, Santiago, Chile}
\altaffiltext{6}{Kavli Institute for Cosmology and Institute of Astronomy, University of Cambridge, Madingley Road, Cambridge CB3 0HA, UK}
\altaffiltext{7}{Department of Physics \& Astronomy, Macquarie University, Sydney, NSW 2109 Australia}
\altaffiltext{8}{Australian Astronomical Observatory, PO Box 296 Epping, NSW 1710 Australia}
\altaffiltext{9}{Department of Physics and Astronomy, University of California, Riverside, CA 92521, USA}
\altaffiltext{10}{Institute for Astronomy, ETH Zurich, 8092 Zurich, Switzerland}
\altaffiltext{\dag}{Hubble Fellow}

\begin{abstract}
Ultra-deep WFC3/IR observations on the HUDF from the HUDF09 program
revealed just one plausible $z\sim10$ candidate UDFj-39546284.
UDFj-39546284 had all the properties expected of a galaxy at $z\sim10$
showing (1) no detection in the deep ACS+WFC3 imaging data blueward of
the F160W band, exhibiting (2) a blue spectral slope redward of the
break, and showing (3) no prominent detection in deep IRAC
observations. The new, similarly deep WFC3/IR HUDF12 F160W
observations over the HUDF09/XDF allow us to further assess this
candidate. These observations show that this candidate, previously
only detected at $\sim$5.9$\sigma$ in a single band, clearly
corresponds to a real source.  It is detected at $\sim$5.3$\sigma$ in
the new $H_{160}$-band data and at $\sim$7.8$\sigma$ in the full
85-orbit $H_{160}$-band stack.  Interestingly, the non-detection of
the source ($<1\sigma$) in the new F140W observations suggests a
higher redshift.  Formally, the best-fit redshift of the source
utilizing all the WFC3+ACS (and IRAC+$K_s$-band) observations is
$11.8\pm0.3$.  However, we consider the $z\sim12$ interpretation
somewhat unlikely, since the source would either need to be
$\sim$20$\times$ more luminous than expected or show very high-EW
Ly$\alpha$ emission (which seems improbable given the extensive
neutral gas prevalent early in the reionization epoch).
Lower-redshift solutions fail if only continuum models are allowed.
Plausible lower-redshift solutions require that the $H_{160}$-band
flux be dominated by line emission such as H$\alpha$ or [OIII] with
extreme EWs.  The tentative detection of line emission at 1.6$\mu$m in
UDFj-39546284 in a companion paper suggests that such emission may
have already been found.
\end{abstract}

\section{Introduction}

As the identification of large numbers of $z\sim8$ galaxies becomes
more routine in deep HST observations (e.g., Bouwens et al.\ 2011b;
Oesch et al.\ 2012b; Bradley et al.\ 2012; Lorenzoni et al.\ 2011),
the high-redshift frontier has clearly moved to $z\sim9$-10.  Only a
small number of $z\sim9$-10 candidates are known to date (Bouwens et
al.\ 2011a, 2013; Oesch et al.\ 2012a; Zheng et al.\ 2012; Coe et
al.\ 2013; Ellis et al.\ 2013).  The quantitative study of $z\sim9$-10
galaxies provides us with our greatest possible leverage for
characterizing the growth rate of galaxies from early times,
clarifying the role that galaxies played in reionizing the universe,
and assessing possible changes in the stellar populations at very low,
even primordial, metallicities.

Of all the $z\sim9$-10 candidates, perhaps the most tantalizing is the
$z\gtrsim10$ candidate UDFj-39546284.  UDFj-39546284 was initially
identified as a promising $z\sim10$ candidate by Bouwens et
al.\ (2011a) making use of the ultra-deep optical and near-IR
observations over the HUDF from the full HUDF09 data set (see also
Oesch et al.\ 2012a).  More recently, UDFj-39546284 was re-examined
using the WFC3/IR observations from the HUDF12 and CANDELS programs by
Ellis et al.\ (2013), and it was found to be undetected in the F140W
band, suggesting that its redshift could be as high as $z\sim11.9$.

In this paper, we perform a detailed reassessment of UDFj-39546284
taking advantage of several additional data sets.  In addition to
utilizing the new ultra-deep WFC3/IR observations from the 128-orbit
HUDF12 (Ellis et al.\ 2013; Koekemoer et al.\ 2013) and CANDELS
(Grogin et al.\ 2011; Koekemoer et al.\ 2011) programs and deep IRAC
observations already considered, we use a deeper reduction of the
optical observations over the HUDF from the XDF dataset (Illingworth
et al.\ 2013) than previously used.  Furthermore, we add new
constraints from a deep $K_s$-band image, add new measurements of
size/structure, and present the source in the context of the expected
$UV$ LF at $z>10$ in a quantitative way.  Finally, we make use of
results from a companion paper (Brammer et al.\ 2013) on deep WFC3/IR
grism spectroscopy of the source to further clarify its nature.

\begin{figure*}
\epsscale{1.16}
\plotone{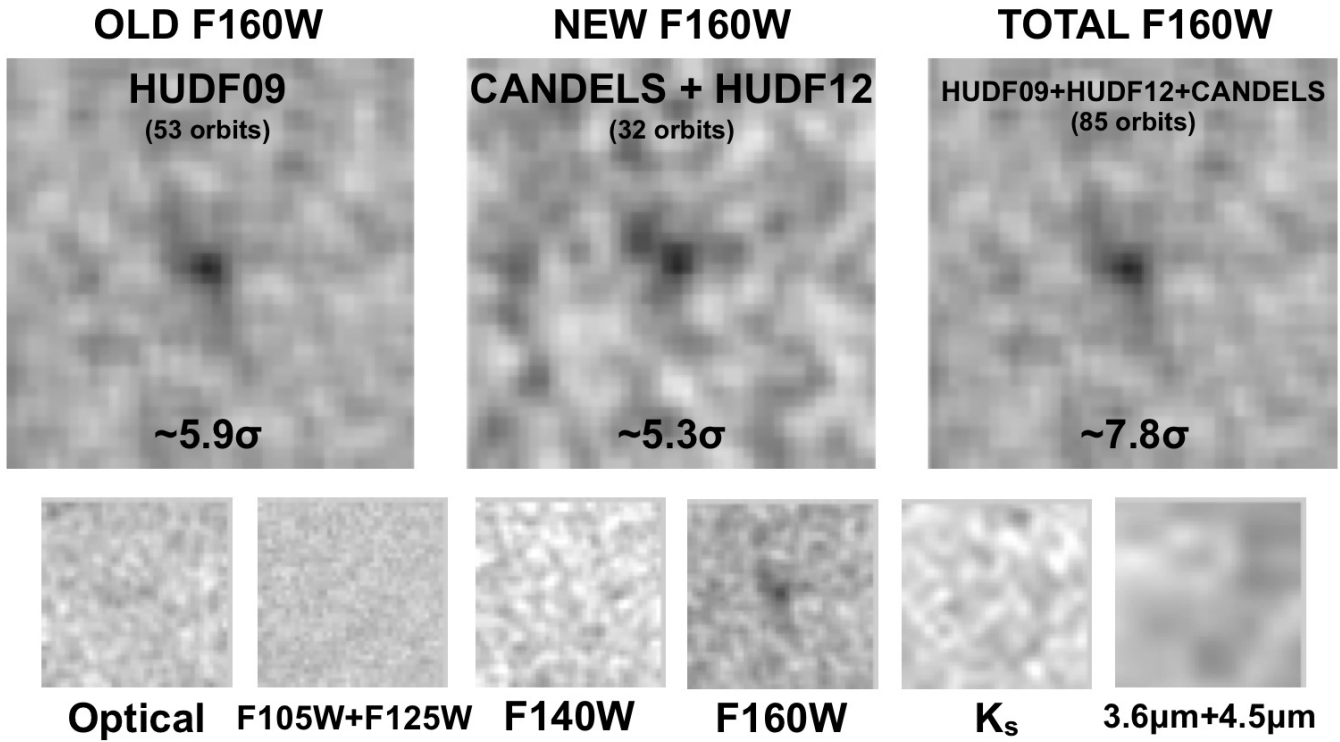}
\caption{$H_{160}$-band imaging observations ($2.4''\times2.4''$) of
  the Bouwens et al.\ (2011a) $z\sim10$ candidate UDFj-39546284 in the
  original 53-orbit HUDF09 observations (\textit{upper-left panel}),
  the new 32-orbit HUDF12+CANDELS observations (\textit{upper-middle
    panel}), and the combined 85-orbit observations
  (\textit{upper-right panel}). Indicated in the lower part of each
  panel is the significance level at which the source is detected in
  the $H_{160}$-band observations (0.5$''$-diameter apertures).  This
  source consists of a bright core embedded in a larger structure
  extending up to 0.4$''$ in radius from the core.  The morphology of
  UDFj-39546284 is similar to a $z=1.61$ [OIII] blob recently
  identified by Brammer et al.\ (2013).  The lower panels show images
  of UDFj-39546284 in the
  optical/$B_{435}V_{606}i_{775}I_{814}z_{850}$, $Y_{105}$+$J_{125}$,
  $JH_{140}$, $H_{160}$, $K_s$, and $3.6\mu$m+$4.5\mu$m
  bands.\label{fig:stamp}}
\end{figure*}

The plan for this paper is as follows.  In \S2, we provide a brief
summary of the observational data.  In \S3, we present the HST
photometry we have for the source and use these observations to
reassess its nature.  Finally, in \S4, we summarize our results.  We
refer to the HST F435W, F606W, F775W, F814W, F850LP, F105W, F125W,
F140W and F160W bands as $B_{435}$, $V_{606}$, $i_{775}$, $I_{814}$,
$z_{850}$, $Y_{105}$, $J_{125}$, $JH_{140}$, and $H_{160}$,
respectively.  Where necessary, we assume $\Omega_0 = 0.3$,
$\Omega_{\Lambda} = 0.7$, $H_0 = 70\,\textrm{km/s/Mpc}$.  All
magnitudes are in the AB system (Oke \& Gunn 1983).

\section{Observational Data and Photometry}

\subsection{Observational Data}

We analyze the full set of HST observations over the HUDF09/XDF,
including data from the 192-orbit HUDF09 program (Bouwens et
al.\ 2011b), CANDELS, the 128-orbit HUDF12 program, and several other
sizeable programs.

255 orbits of HST WFC3/IR observations are now available over the
HUDF09/XDF, including $\sim$100, $\sim$40, 30, and $\sim$85 orbits in
the $Y_{105}$, $J_{125}$, $JH_{140}$, and $H_{160}$ bands,
respectively.  The biggest gains over the HUDF09 program came in the
$Y_{105}$ and $JH_{140}$ bands.  We reduced these observations in a
similar manner as the original WFC3/IR data from the HUDF09 program
(Bouwens et al.\ 2011b).  Special care was taken to keep our
reductions of the new CANDELS and HUDF12 observations separate from
those of the original HUDF09 data, to enable us to evaluate the
reality of sources from the original observations.

In addition, we now have new reductions of the optical observations
over the HUDF from the XDF project that are $\sim$0.1-0.2 mag deeper
than the original Beckwith et al.\ (2006) HUDF reductions due to our
inclusion of all other HST data sets taken over the HUDF for the past
10 years, including the recent optical/ACS $I_{814}$ data.

To obtain photometric constraints on UDFj-39546284 redward of the
$H_{160}$-band, we utilize the deep 120-hour Spitzer/IRAC (Fazio et
al.\ 2004) observations in the [3.6] and [4.5] channels from the
original GOODS IRAC program and the 262-hour IRAC Ultradeep Field
program (IUDF10: PI: Labb{\'e}).  These observations reach to 27.1 mag
and 26.8 mag in the 3.6$\mu$m and 4.5$\mu$m bands, respectively
(3$\sigma$: Labb{\'e} et al.\ 2012).  We also utilize the very deep
$K_s$-band observations over the HUDF (26.5 mag: $5\sigma$), including
data from VLT/HAWK-I (program 186.A-0898, PI A. Fontana), VLT/ISAAC
(program 73.A-0764 PI I. Labb{\'e} and 168.A-0485 PI C. Cesarsky), and
PANIC (PI I. Labb{\'e}).  In coadding the $K_{s}$-band observations,
individual frames are weighted by the inverse variance expected for a
point source (Labb{\'e} et al.\ 2003).  Table~1 provides detailed
information on the various $K_s$-band observations used for our deep
reduction.

In summary, in addition to the HUDF09/HUDF12 WFC3/IR and the IUDF10
IRAC datasets (also used by Ellis et al. 2013, though Ellis et
al. 2013 appear not to have accounted for the modest variations in the
effective depth of the IRAC observations due to varying contamination
from neighboring foreground sources), we utilize the deeper XDF
optical/ACS dataset on the HUDF, deep optical/ACS $I_{814}$ and deep
$K_s$-band observations for our detailed study of UDFj-39546284.

\begin{figure}
\epsscale{1.16}
 \plotone{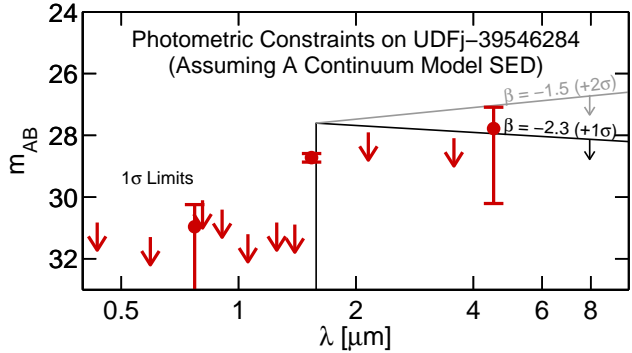}
 \caption{Photometric constraints on the SED of UDFj-39546284.  The
   solid red points and error bar show the flux measurements, with
   $1\sigma$ uncertainties, and $1\sigma$ upper limits.  The source is
   only detected ($>7\sigma$) in the $H_{160}$ band.  The source shows
   a very large decrement between the $H_{160}$ band and the
   $JH_{140}$ band ($>$2.2 mag).  The dark and light gray lines show
   the $1\sigma$ and $2\sigma$ upper limits, respectively, that can be
   set on the spectral slope $\beta$ redward of the break at
   $\sim$1.6$\mu$m, assuming a continuum-model SED.  The limits
   redward of the $H_{160}$-band show that UDFj-39546284 is blue in
   color and not red.\label{fig:beta}}
\end{figure}

\subsection{Methodology for Doing Photometry}

We obtain flux measurements on the HST observations by running
SExtractor (Bertin \& Arnouts 1996) in dual-image mode, taking the
detection image to be the $H_{160}$-band and using the PSF-matched
images for photometry.  Colors are measured in small-scalable
apertures (Kron [1980] factor of 1.2) and corrected to total by
comparing the $H_{160}$-band flux in a larger-scalable aperture (Kron
factor of 2.5) to that in the smaller-scalable aperture and then
applying a correction to account for light on the wings of the PSF
based on the tabulated encircled-energy distribution.

The $K_s$-band flux measurement was performed in a $0.65''$-diameter
circular aperture and corrected to match our HST photometry by
comparing the $H_{160}$-band flux ($0.65''$-diameter aperture, after
PSF-correction) to that found in our baseline scalable apertures.

IRAC photometry was performed utilizing software to model the light
profiles of neighboring sources so that this light could be subtracted
(Labb{\'e} et al.\ 2006, 2010a, 2010b, 2012).  Clean photometry of the
source is then performed (1.8$"$-diameter circular apertures).  A
factor of $\sim$2.2 correction is made to the measured fluxes to
account for light on the wings of the IRAC PSF.

\section{Results}

\subsection{Photometric Constraints on UDFj-39546284}

\begin{figure}
\epsscale{1.16}
\plotone{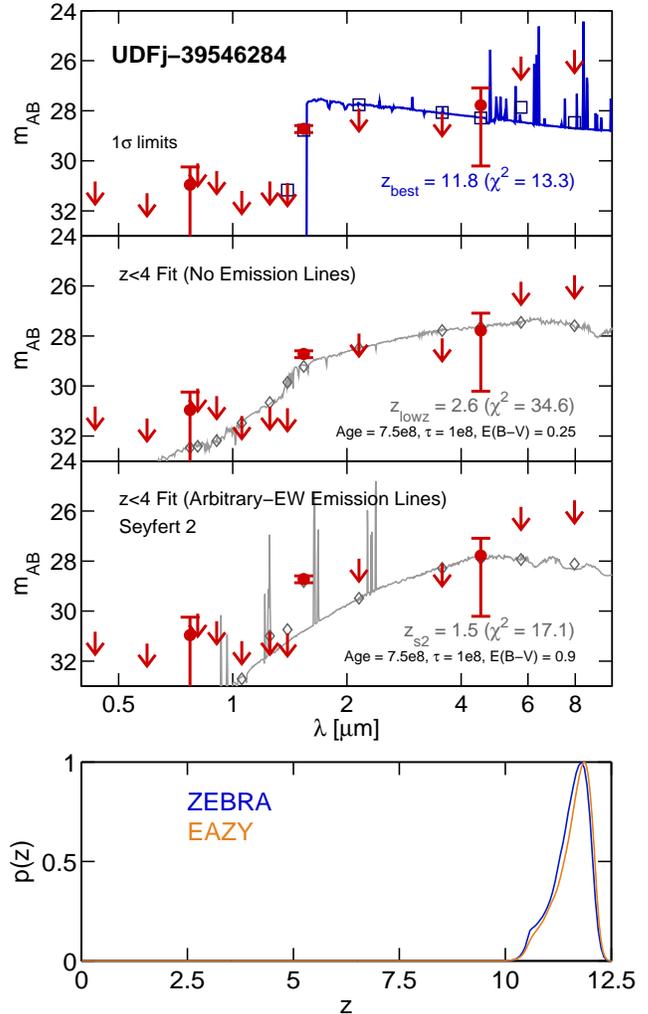}
 \caption{The flux constraints shown in the top three panels are as in
   Figure~\ref{fig:beta}. The upper panel shows the best-fit SED at
   high redshift with $z_{phot}=11.8\pm0.3$ for a source without
   Ly$\alpha$ emission. The upper-middle panel shows the poor, but
   best low-redshift fit ($z_{lowz}=2.6$, modest reddening
   $E(B-V)=0.25$) to the available photometry without invoking
   emission lines.  The lower-middle panel shows the best low-redshift
   solution, allowing for the inclusion of arbitrary-EW emission lines
   from an AGN (Seyfert 2), with $z_{s2}=1.5$, reddening $E(B-V)=0.9$,
   and H$\alpha$ EW$_{0}\sim5000$\AA.  None of these ``solutions'' is
   especially likely (see text).  The lowest panel shows the derived
   redshift likelihood distributions using the ZEBRA and EAZY
   photometric redshift codes.\label{fig:JdropSED}}
\end{figure}

The photometry we derive for UDFj-39546284 is presented in
Table~\ref{tab:photometry} and Figure~\ref{fig:beta}.  UDFj-39546284
again shows a very significant detection in the $H_{160}$-band and no
significant detection in any other band.  The fact that the source is
detected in the new $H_{160}$-band observations at $5.3\sigma$
(0.5$''$-diameter aperture) and 7.8$\sigma$ (0.5$''$-diameter
aperture) in the total $H_{160}$-band stack demonstrates that this is
definitely a real galaxy (Figure~\ref{fig:stamp}).

The present color measurements are consistent with those from Ellis et
al.\ (2013), but our total $H_{160}$-band magnitude is $\sim$0.6 mag
brighter than the $0.5''$-diameter aperture-magnitude measurement from
Ellis et al.\ (2013).  This is not unsurprising given the significant
spatial extension of UDFj-39546284, our use of larger scalable
apertures (more appropriate for this source), and our correction to
total magnitudes.

The optical and near-IR observations blueward of the $H_{160}$-band
are very deep and indicate a sharp fall-off in the spectrum at
$<$1.6$\mu$m (Figure~\ref{fig:beta}).  The amplitude of the flux
decrement from the $H_{160}$-band flux measurement is a substantial
$>$2.8 mag to a coadded $Y_{105}J_{125}JH_{140}$ bandpass, $>$3.3 mag
to a coadded
$B_{435}V_{606}i_{775}I_{814}z_{850}Y_{105}J_{125}JH_{140}$ bandpass,
and $>$2.2 mag to the $JH_{140}$-band.  (The flux constraints from
multiple bands were combined assuming a flat-spectrum (F$\nu$)
source.)  Redward of the $H_{160}$-band, the IRAC and $K_s$-band
observations are less deep, but place strong constraints on the
general shape of the SED.

The existence of a significant $\sim$8$\sigma$ detection of
UDFj-39546284 in the $H_{160}$-band, a strong break in the spectrum
blueward of the $H_{160}$-band, and no prominent detection of the
source redward of the $H_{160}$-band is suggestive of a $z>10$ galaxy.
Use of the photometric redshift code ZEBRA (Feldmann et al.\ 2006)
yields a formal redshift estimate of $11.8\pm0.3$ for UDFj-39546284
(Figure~\ref{fig:JdropSED}).  We find a similar result with the EAZY
photometric redshift code (Brammer et al.\ 2008).  The present
estimates are somewhat lower than the Ellis et al.\ (2013) $z=11.9$
estimate, likely due to our additional constraint on the luminosity of
UDFj-39546284 from the deep $K_s$-band data.

As in Oesch et al.\ (2012a) and Ellis et al.\ (2013), attempts to fit
the source with a lower-redshift galaxy SED are not particularly
successful.  The best low-redshift fit, without a substantial
emission-line contribution, is an evolved galaxy at $z_{lowz}=2.6$.
However, the high measured $\chi^2$ value of this fit
($\chi^2_{lowz}=34.6$) compared to the best-fit solution at $z=11.8$
($\chi^2=13.3$) makes this conventional low-redshift solution
untenable (but see below).

The new photometry also allows us to set useful constraints on the
shape of the spectrum redward of the break.  We fit the SED with a
power law $f_{\lambda}\propto\,f_0\,\lambda^{\beta}$, leaving the
redshift, luminosity, and $\beta$ as free parameters.  We find
$1\sigma$ and $2\sigma$ upper limits of $-$2.3 and $-$1.5,
respectively, for $\beta$.  These upper limits correspond to the
maximum $\beta$'s where $\Delta\chi^2=\chi^2-\chi_{min}^2=1$ and 4,
respectively.  This demonstrates quite definitively that UDFj-39546284
is blue redward of the $H_{160}$-band and cannot be well fit by an old
or dusty low-redshift SED.

\begin{figure}
\epsscale{1.16}
 \plotone{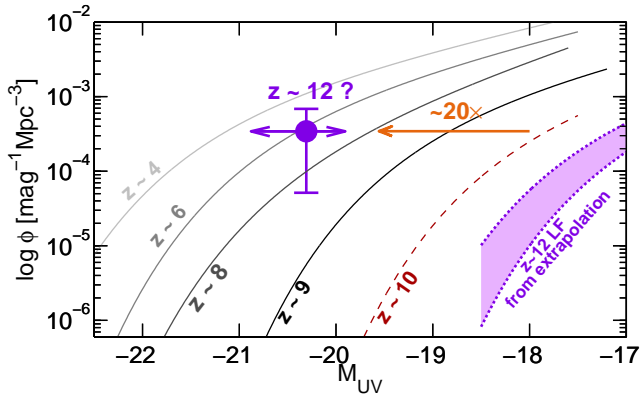}
 \caption{UDFj-39546284 is extremely luminous, if it is a $z\sim12$
   galaxy.  Shown is the constraint on the $z\sim12$ LF we would
   derive, if UDFj-39546284 were genuinely a $z\sim12$ star-forming
   galaxy.  The effect of the photometric-redshift uncertainties
   $z=11.8\pm0.3$ on the inferred luminosity are indicated by the
   horizontal arrows.  For context, the $z\sim4$-10 LFs presented in
   Bouwens et al.\ (2007), Oesch et al.\ (2012a), and Oesch et
   al.\ (2013) are also shown.  UDFj-39546284 is much more luminous,
   by a factor of $\sim$10, than comparably-prevalent galaxies at
   $z\sim10$ (i.e., with the same volume density).  The luminosity of
   UDFj-39546284 is even more anomalous compared to the LF
   extrapolated to $z\sim12$ using $z\geq6$ and $z\geq8$ LF trends
   (Oesch et al.\ 2012a, 2013).  In that case UDFj-39546284 would be
   at least $\sim$20$\times$ more luminous.  Given the uniform rate of
   evolution in the $UV$ LF to early times, it is very unlikely that
   we would find such a luminous galaxy at
   $z\sim12$.\label{fig:LFconstraint}}
\end{figure}

\subsection{Size and Structural Properties of UDFj-39546284}

UDFj-39546284 consists of a compact core, embedded in a more extended
structure.  The features to the upper left and lower right of the
source (Figure~\ref{fig:stamp}) appear to extend some $\sim$0.4$''$ in
radius from the source (see also Ono et al.\ 2013).  Using SExtractor,
we measure a half-light radius of $\sim$0.17$"$ for UDFj-39546284 in the
deeper observations.  This is larger than the $\sim$0.13$"$-diameter
half-light radius for the PSF.  Correcting for the PSF, the half-light
radius for this candidate is 0.13$''$.

If we interpret this as a $z\sim12$ source, the implied $\sim$0.5 kpc
(physical) half-light radius for the source is consistent with
expectations what one would applying a $(1+z)^{-1}$ size scaling to
the $z\sim7$-8 galaxy samples studied by Oesch et al.\ (2010) where
the mean size for comparable-luminosity sources is 0.8 kpc (physical:
see also Ono et al.\ 2013).  A $\sim(1+z)^{-1}$ size scaling has been
found to describe the evolution of star-forming galaxies over a wide
range in redshift (e.g., Buitrago et al.\ 2008; Oesch et al.\ 2010).
This source is also potentially consistent with expectations if we
interpret this as a $z\sim2$ galaxy.  The measured half-light radius
translates into a physical size of $\sim$1.1 kpc, at the top end of
the range expected for extreme emission-line or star-forming galaxies
at $z\sim2$ (van$\,$der$\,$Wel et al.\ 2012), after correcting for
typical $r\propto\,L^{0.3}$ luminosity dependencies (e.g., de$\,$Jong
\& Lacey 2000).

\subsection{Difficulty with $z\sim12$ Interpretation: Inferred $UV$ Luminosity Is $\sim$20$\times$ Too Large?}

While simply identifying UDFj-39546284 as a $z\sim12$ galaxy would
seem appropriate (see also Ellis et al.\ 2013), it becomes problematic
when both the apparent UV luminosity of this source and the total
search volume in which the source was found are taken into
consideration.  If the source is at $z\sim12$, its intrinsic
luminosity would be $-$20.3 ($\sim$0.5$\times$ the luminosity of a
$z=3$ $L^*$ galaxy: Steidel et al.\ 1999). This is $\sim$4$\times$
higher than what Bouwens et al. (2011a) and Oesch et al.  (2012a)
inferred if the source was at $z\sim10.4$ (which seemed plausible
before the $JH_{140}$ constraint was available).  The much higher
luminosity follows from the greater luminosity distance for
UDFj-39546284 (factor of $\sim$1.4 change) and the fact that the
source is only seen in the reddest one-third of the $H_{160}$-band
(factor of 3 change).

To put this unusually high luminosity in context, we calculate the
approximate volume in which we could have found the source.  Utilizing
the same techniques as in Oesch et al.\ (2013), we estimate a total
search volume of $\sim$3$\times$10$^3$$\,$Mpc$^3$ (comoving) for
UDFj-39546284 in the combined HUDF09/HUDF12/XDF dataset.

The calculated selection volume and observed $UV$ luminosity allow us
to derive an approximate LF for star-forming galaxies at $z=12$,
assuming of course that UDFj-39546284 is indeed at $z=12$.  The result
is shown in Figure~\ref{fig:LFconstraint} and is unique to this
analysis.  For context, the LFs inferred for star-forming galaxies at
$z\sim4$-10 from Bouwens et al.\ (2007), Oesch et al.\ (2012b), and
Oesch et al.\ (2013) are also presented.  It is clear that
UDFj-39546284 would be $\sim$10$\times$ more luminous than
similarly-prevalent galaxies at $z\sim10$ and $\sim$20$\times$ more
luminous than similarly-prevalent star-forming galaxies at $z\sim12$,
extrapolating lower-redshift LF trends to $z>10$.

The evolution of the $UV$ LF at early times, i.e., from $z\sim10$ to
$z\sim4$, is sufficiently uniform that the discovery of a $z\sim12$
galaxy that is $\sim$20$\times$ more luminous than expected over such
a small area is implausible and strongly argues for another
explanation.\footnote{We remark that gravitational lensing by a
  foreground source does not seem like a workable explanation for the
  high luminosity of UDFj-39546284, given the lack of a plausible
  foreground lens.}

\begin{deluxetable*}{cc|cccc}
\tablecaption{Photometry of UDFj-39546284 and Observations Utilized in Constructing A Deep $K_s$-band Image of the HUDF09/HUDF12/XDF.\label{tab:photometry}}
\tablewidth{10cm}
\tablehead{\multicolumn{2}{c|}{Photometry of UDFj-39546284} & \multicolumn{4}{c}{$K_s$-band Observations}\\
 & & & $t_{exp}$ & Depth & Seeing\\
Quantity & Measurement & Instrument & [hrs] & ($5\sigma$) & FWHM [``] }
\startdata
RA & 03:32:39.54 & VLT/HAWK-I & 28.4 & 26.1 & 0.36 \\
DEC & $-$27:46:28.4 & VLT/ISAAC & 24.2 & 25.8 & 0.35 \\
$B_{435}$ & $-1.0\pm1.7$ & PANIC & 23.6 & 25.5 & 0.33 \\
$V_{606}$ & $0.2\pm1.1$ & \textit{ALL} & 76.2 & 26.5 & 0.35 \\
$i_{775}$ & $1.5\pm1.4$ \\
$I_{814}$ & $-2.9\pm3.3$\\
$z_{850}$ & $1.2\pm2.5$ \\
$Y_{105}$ & $-0.8\pm1.2$ \\
$J_{125}$ & $-3.9\pm1.7$ \\
$JH_{140}$ & $-0.5\pm1.6$ \\
$H_{160}$ & $11.8\pm1.5$ \\
 & (28.7$\pm$0.2 mag)\\
$K_s$ & $-16\pm25$ \\
$3.6\mu$m & $4\pm21$ \\
$4.5\mu$m & $28\pm25$ \\
$5.8\mu$m & $-36\pm168$ \\
$8.0\mu$m & $-136\pm215$
\enddata
\tablecomments{Fluxes (corrected to total: \S2.2) are given in nJy.}
\end{deluxetable*} 

\subsection{Emission-Line-Dominated Galaxy?}

The properties of UDFj-39546284 are puzzling and difficult to explain
as either a low or high-redshift source if the bulk of the
$H_{160}$-band flux is from continuum star light.

However, we can avoid this difficulty if the $H_{160}$-band light
predominantly arises from line emission (see also Ellis et al.\ 2013).
For example, in the lower-middle panel of Figure~\ref{fig:JdropSED},
we show one possible, though somewhat extreme example, where we allow
for arbitrary-EW line emission from an AGN.  The dust extinction is
high ($E(B-V)=0.9$), and the rest-frame EW of $H_\alpha$ is large
($\sim$5000\AA).  This fit has a $\chi^2$($=17.1$) more similar to the
$z\sim12$ solution, but for a redshift $z\sim1.5$.  This is ad-hoc,
but demonstrates what is needed.

Support for line emission contributing substantially to the flux in
the $H_{160}$-band comes from the recent analysis of the deep WFC3/IR
grism observations of UDFj-39546284 in a companion paper by Brammer et
al.\ (2013).  They find evidence of an emission-line feature at
1.6$\mu$m that could provide most or all of the observed
$H_{160}$-band flux for UDFj-39546284. The existence of such a feature
is not unexpected given the difficulty in modelling the source as a
pure-continuum galaxy at $z\sim12$ (because of its luminosity) or
$z\sim2$-3 ($\chi^2=34.6$: Figure~\ref{fig:JdropSED}).

Given the plausiblity of line emission playing a role, the question
arises as to the nature of the line emission.  Brammer et al.\ (2013)
argue that the emission-line contribution would be from an extreme
emission-line galaxy (EELG), notably the [OIII]$\lambda$4959+5007
doublet at $z\sim2.2$.  Such galaxies have been found in wide-area
grism and imaging surveys with WFC3/IR (van$\,$der$\,$Wel et
al.\ 2011; Atek et al.\ 2011; Brammer et al.\ 2012).  Even more
extreme examples are needed in the case of UDFj-39546284.  Ellis et
al.\ (2013) similarly suggested the source might be an EELG at
$z\sim2.4$, but could not explain how such a source could produce the
observed spectral break.  Brammer et al.\ (2013) describe the
discovery of an EELG at $z\sim1.6$ with an extremely-high [OIII] EW
and relatively-red UV colors that would come close to satisfying the
constraints if that source were at $z\sim2.2$.

Alternatively the emission-line flux could be from Ly$\alpha$.  EWs of
$\sim$200$\,$\AA$~$are seen in star-forming galaxies in the $z\sim4$-6
universe (e.g., Stark et al.\ 2010) and would cause the source to be
brighter by a factor of $\sim$4, resulting in a much more plausible
intrinsic luminosity, i.e., $\sim-18.8\,$mag.  However, even with this
luminosity, the source would still be at least 5$\times$ more luminous
than one would expect for a comparably-prevalent $z\sim12$ galaxy
(i.e., with the same volume density).  Attributing the emission to
Ly$\alpha$ also seems implausible given the large amounts of neutral
hydrogen expected in the $z>7$ universe that would resonantly scatter
Ly$\alpha$ photons.  Ly$\alpha$ emission is found to be rare in
$z\gtrsim7$-8 galaxies, presumably due to an increasingly neutral IGM
(Ono et al.\ 2012; Schenker et al.\ 2012; Pentericci et al.\ 2011;
Caruana et al.\ 2012).

The existence of line emission is a plausible, though not proven,
solution to the mystery regarding UDFj-39546284, with the evidence
weighted towards a low-redshift solution with an extremely strong
[OIII] feature.

\section{Summary}

We utilize the deeper near-IR observations available over the
HUDF09/HUDF12/XDF from the HUDF09, HUDF12 and CANDELS programs to
investigate the nature of the $z\sim10$ galaxy candidate UDFj-39546284
(Bouwens et al.\ 2011a; Oesch et al.\ 2012a).  Using the
$H_{160}$-band observations from the combined HUDF12 and CANDELS
programs, we find a 5.3$\sigma$ detection at the position of
UDFj-39546284, definitively demonstrating that the candidate is real
(see also Ellis et al.\ 2013; Ono et al.\ 2013).  UDFj-39546284 is
detected at $\sim$7.8$\sigma$ in the full 85-orbit $H_{160}$-band
observations.

Making use of deeper ACS+WFC3/IR XDF observations we demonstrate that
UDFj-39546284 exhibits a substantial break in the SED between the
$H_{160}$-band and bluer bands: $\gtrsim2.2$ mag to the $JH_{140}$
band and $\gtrsim3.3$ mag to a combined
$B_{435}V_{606}i_{775}I_{814}z_{850}Y_{105}J_{125}JH_{140}$ band.
Using the deep IRAC and $K_s$-band observations, we find that
UDFj-39546284 has $1\sigma$ and $2\sigma$ upper limits of $-2.3$ and
$-1.5$, respectively, for the $UV$ slope $\beta$ showing quite clearly
that UDFj-39546284 is not a red $z\sim1$-3 galaxy
(Figure~\ref{fig:beta}).

The sharp break in the SED of UDFj-39546284 and blue spectral slope
redward of the break is suggestive of a $z\sim10$-12 galaxy.  The
best-fit photometric redshift for UDFj-39546284 using all data,
including deep IRAC and $K_s$-band constraints, is $z=11.8\pm0.3$.

However, interpreting the source as a $z=11.8\pm0.3$ star-forming
galaxy is problematic.  The $UV$ luminosity inferred for the source if
it were at $z\sim12$ is extremely high, $\sim$20$\times$ brighter than
expected for similarly-prevalent sources at $z\sim12$ (extrapolating
current LF trends).

In light of the uniform evolution seen in the $UV$ LF at early times,
it seems implausible that UDFj-39546284 actually corresponds to a
$z\sim12$ galaxy unless its $H_{160}$-band flux is substantially
boosted by Ly$\alpha$ emission.  However, this possibility is unlikely
given the huge amounts of neutral hydrogen almost certainly present in
the $z\sim12$ universe.

Given the tentative detection of an emission line in UDFj-39546284 by
Brammer et al. (2013), the most probable interpretation for
UDFj-39546284 may be that it corresponds to a rare EELG at $z\sim2.2$
with an observed [OIII] EW $>$10$^4$\AA.  An example of such an EELG
is presented by Brammer et al.\ (2013).  Such a high-EW source is even
more extreme than other recently-identified EELGs (e.g.,
van$\,$der$\,$Wel et al.\ 2011).

While UDFj-39546284 is not at $z\sim10$, and probably not at
$z\sim12$, the outcome is comparably interesting, and exemplifies the
challenges of exploring the limits of the high-redshift universe with
current telescopes as we await JWST.

\acknowledgements

We acknowledge the support of NASA grant NAG5-7697 and NASA grant
HST-GO-11563, ERC grant HIGHZ \#227749, and a NWO vrij competitie
grant.  PO acknowledges support from NASA through a Hubble Fellowship
grant \#51278.01 awarded by STScI.

\end{document}